\documentclass[11pt,twoside]{article}
\usepackage{./asp2014}
\usepackage{graphicx}
\usepackage{natbib}

\resetcounters

\begin{document}

\title{Gas Density Across the Universe}
\author{Adam K. Leroy,$^1$ Alberto Bolatto,$^2$ Amanda Kepley,$^3$ David Meier,$^4$ and Tony Wong$^5$}
\affil{$^1$Ohio State University, Department of Astronomy, Columbus, OH 43210; \email{leroy.42@osu.edu} }
\affil{$^2$Department of Astronomy, University of Maryland, College Park, MD 20742}
\affil{$^3$National Radio Astronomy Observatory, 520 Edgemont Road, Charlottesville, VA 22903-2475}
\affil{$^4$New Mexico Institute of Mining and Technology, 801 Leroy Place, Socorro, NM 87801}
\affil{$^5$Department of Astronomy, University of Illinois, Urbana, IL 61801}

\paperauthor{Adam Leroy}{leroy.42@osu.edu}{}{Ohio State University}{Department of Astronomy}{Columbus}{OH}{43210}{USA}

\begin{abstract}
Gas density is widely believed to play a governing role in star formation. However, the exact role of density in setting the star formation rate remains debated. We also lack a general theory that explains how the gas density distribution in galaxies is set. The primary factor preventing the resolution of these issues is the limited number of observations of the gas density distribution across diverse environments. Centimeter- and millimeter-wave spectroscopy offer the most promising way forward in this field, but the key density-sensitive transitions are faint compared to the capabilities of current telescopes. In this chapter, we describe how a next-generation Very Large Array (ngVLA) represents the natural next step forward in this sensitivity-limited field. Such a facility would provide a crucial link between the ``Milky Way'' and ``Extragalactic'' views of star formation and dramatically advance our understanding of the drive and role of gas density in galaxies, building on current results from ALMA, NOEMA, the Green Bank Telescope, and other current facilities working in this area.
\end{abstract}

\section{Is Density Destiny in Star Formation?}

Gas density and star formation are closely linked. Observationally, several lines of evidence show a strong relationship between gas density and star formation. Within the Milky Way, the amount of dense gas is the best predictor of the current rate of star formation in nearby clouds \citep[e.g.,][]{EVANS14}. Specifically, star formation appears associated with dense, filamentary substructure within these clouds \citep[see review by][]{ANDRE14}. On larger scales, integrated measurements of entire galaxies demonstrate a linear correlation between the rate of recent star formation and the amount of dense molecular gas traced by high critical density emission lines, especially HCN and HCO$^+$ \citep[e.g.,][]{GAO04}.

This close association makes sense, given that gas density plays a key role in nearly every theory of star formation. But so far our observations of dense gas have been severely limited by the sensitivity of our telescopes. The key lines to observe dense gas are more than an order of magnitude fainter than the workhorse CO lines. In the absence of the kind of sensitive, resolved observations of dense gas that will be made by the ngVLA, the exact role played by density in star formation remains relatively unconstrained.

Recent theories of star formation disagree about whether the mean density, only the fraction of dense gas, or the distribution of densities is the key factor setting the rate of star formation.  In the first category, the governing timescale to convert gas into stars is the gravitational free-fall time at some characteristic scale. Then this timescale, and so the normalized rate of star formation, is set by the mean density of the ISM at that scale \citep[e.g.,][]{KRUMHOLZ12,ELMEGREEN18}. In a theory like this, for example, the mean density of molecular cloud will govern its ability to form stars. In a second class of models,  gas above some threshold density forms stars in a semi-universal way \citep[e.g.,][]{GAO04,LADA12}. Such models focus on the formation of dense gas, however it happens, as the key limiting step to form stars. A third category of theories considered the distribution of densities within a cloud \citep[often treated as turbulent, isothermal gas following, e.g.,][]{PADOAN02,KRUMHOLZ05} to play the key role in determining its ability to form stars \citep[e.g.,][]{HENNEBELLE11,FEDERRATH12}.

We also know that the density of gas changes in response to the environment within a galaxy. For example, major mergers (such as the local ultra-luminous infrared galaxy population) appear to have a much higher mean gas density than normal spiral galaxies \citep[][]{GAO04,GARCIABURILLO12}. The central parts of disk galaxies --- including the Milky Way's own Central Molecular Zone --- show higher gas densities than the more quiescent outer parts of disks \citep[e.g.,][]{LONGMORE13, USERO15,CHEN15,BIGIEL16,GALLAGHER18}. Indeed, gas density also appears to change systematically across spiral galaxies, with denser gas in high surface density, gas rich environments. Though there are suggestions of a link to interstellar gas pressure \citep[e.g.,][]{HELFER97}, we lack a predictive theory that explains how the structure and properties of a galaxy set its gas density distribution.

The central role of density in star formation theory, the close association of molecular gas density and star formation, and the observation that gas density changes in response to environment prompts a series of questions:

\medskip

\noindent {\em 1) What is the distribution of gas volume densities in the molecular interstellar medium?} 

\smallskip 

\noindent In theoretical work, this is often discussed as the density probability distribution function (PDF). Knowledge of the density PDF is central to understand star formation (see above), feedback \citep[e.g.,][]{KRUMHOLZ16}, and emission from molecular lines \citep[e.g.,][]{LEROY17a}. Based on theoretical work, a lognormal, power law, or some hybrid distribution is often assumed \citep[see][]{PADOAN02,KRUMHOLZ05}. However, observational constraints on this quantity remain weak. The best observational measurements currently consist of column density measurements in local clouds \citep[e.g.,][]{ABREUVICENTE15}.  However, as discussed above, observations clearly indicate that the mean gas density changes across the universe, and resolved observations of local clouds offer a window into only a very narrow subset of the physical conditions 
found across the universe.

\medskip

\noindent {\em 2) How does gas density relate to star formation in different environments?} 

\smallskip

\noindent As described above, observations suggest a strong link between gas density and star formation. However, the quantitative relationship between the gas density distribution and the formation of stars and clusters remains weakly constrained by observations. The strongest possible test of many current theories of star formation in molecular clouds would be detailed observations of the density distribution, dynamical state, and star formation rate of molecular clouds in a wide range of environments and evolutionary states. These observations would clearly resolve whether there some threshold density above which star formation proceeds in a universal way and whether the gravitational free-fall time is indeed the governing timescale for star formation over many scales.

\medskip

\noindent {\em 3) How does gas density relate to galactic structure and dynamics?} 

\smallskip

\noindent The gas density distribution clearly changes across the universe, with denser gas found in environments with deep potential wells and strong compressive effects (e.g., galaxy mergers, spiral arms, gas flows along bars). Our quantitative knowledge of how gas density responds to environment, and vice versa, is in its infancy, however.

\section{Sensitive cm-and mm-Wave Spectroscopy as the Path Forward}

\begin{figure}[t!]
\includegraphics[width=\textwidth]{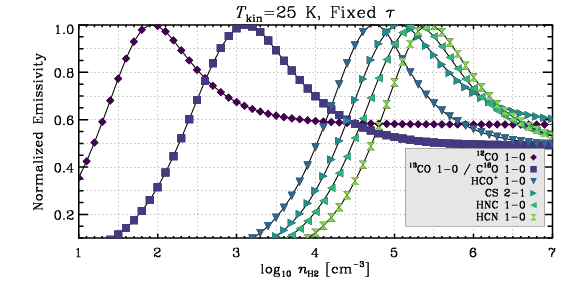}
\caption{{\bf Emissivity vs. Density for Some cm- and mm-Wave Lines.} Emissivity (intensity per column density) as a function of collider volume density for a subset of lines in the ngVLA high frequency window \citep[following][]{LEROY17a}. The ngVLA covers low-$J$ lines (which are relatively less sensitive to excitation effects) that emit effectively under a wide range of gas densities. Combing these lines offers the prospect to constrain the density distribution point by point across galaxies. The sensitivity, bandwidth (allowing heavy multiplexing), and resolution of the ngVLA are unmatched for this kind of work, giving it the prospect to act as an unparalleled density mapping machine.}\label{fig:emis}
\end{figure}

Thee preceding questions remain major open issues because observing molecular gas density is incredibly challenging. The densest regions of molecular clouds are typically, $\sim 0.1{-}1$~pc in scale \citep[see][]{LADA03,ANDRE14}. In Milky Way molecular clouds, one can achieve the resolution to pick out individual dense substructures, and so reconstruct the density distribution from imaging. Observations in the Milky Way, and the Solar Neighborhood, however, probe only a limited subset of the environments found across the universe.

Even with the ngVLA, is simply infeasible to image gas at $0.1{-}1$~pc resolution beyond the handful of nearest galaxies. As a result, cm- and mm-wave spectroscopy is our best tool to study the density distribution in more distant systems. The cm- to mm-wave part of the spectrum contains rotational transitions from a host of interstellar molecules. These transitions, in turn, have a wide range of critical densities. The emissivity of gas in any particular molecular transition falls off quickly with decreasing density below the effective critical density. By measuring the relative intensities of lines with a wide range of effective critical densities, one can thus probe a wide range of different gas densities. We can use these observations to constrain the distribution of mass as a function of gas volume density within a beam (i.e., to constrain the density PDF in the beam). This technique works without any need to resolve individual sub-parsec scale dense substructures because the spectroscopic tracers already preferentially trace gas near the effective critical density of the line.

This approach holds huge promise: the ability to measure the distribution of densities in each cloud, or even each moderate size region, across a galaxy will allow us to concretely link this key driver for star formation on cloud scales (density) to galactic environment. This push towards measuring physical conditions in the cold gas across galaxies is a major thrust of star formation studies in $\sim 2020$ and is sure to remain a key area moving in to the era of the ngVLA.

\section{The Need for the ngVLA}

\begin{figure}[t!]
\includegraphics[width=0.45\textwidth]{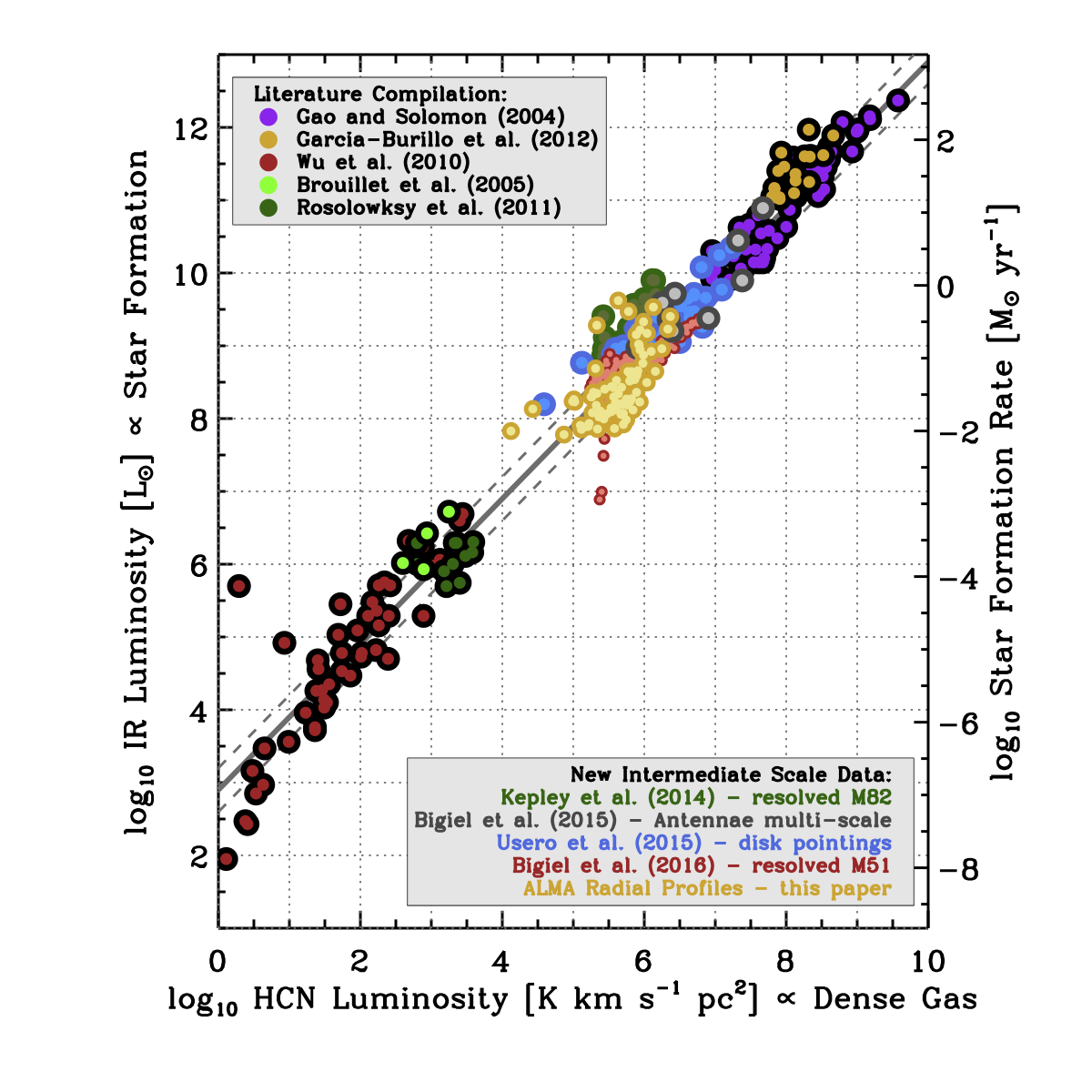}
\includegraphics[width=0.45\textwidth]{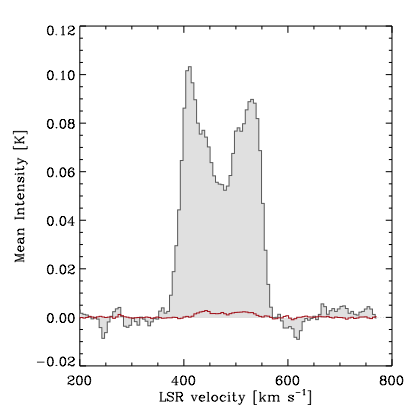}
\caption{{\bf The Promise and Challenge of Mapping Dense Gas.}\label{fig:sketch} (\textit{Left:}) From \citet{GALLAGHER18}, the correlation between recent star formation and dense gas seen over $15$ years of HCN studies \citep[following][]{GAO04,WU05,WU10}. Across decades in scale, tracers of high density gas correlate well with star formation, though the detailed physics have yet to be well-constrained by observations. (\textit{Right:}) The integrated HCN spectrum of M51 \citep[from][and see \citealt{CHEN15}]{BIGIEL16} plotted in red over the integrated CO spectrum of the galaxy in gray. Even the brightest ``dense gas tracers'' are $\sim 10{-}100$ times fainter than CO emission. As a result, the sort of density-probing spectroscopy discussed in this chapter strains the capabilities of current facilities. The sensitivity of the ngVLA will greatly surpass the sensitivity of these facilities, enabling us to map dense gas tracers at high resolution across the local universe ($\sim$ Virgo).}
\end{figure}

The main obstacle to deploying this approach widely is the faintness of the high critical density transitions available in the cm- and mm-wave window. The brightest transitions in this window include rotational transitions of HCN and HCO$^+$. These molecules are often $\sim 10^8$ times less abundant than H$_2$ and $\sim 10^4$ times less abundant than CO \citep[e.g.,][]{MARTIN06}, and the dense gas often makes up only $\sim 1{-}10\%$ of a cold cloud by mass \citep[e.g.,][]{LADA10}. As a result, even the brightest high critical density lines (``dense gas tracers'') are extremely faint compared to the standard molecular gas-tracing CO lines. Line ratios between HCN (the brightest dense gas tracer) and CO seldom exceed $0.1$ across a large part of a galaxy and typical values in galaxy disks are $\sim 0.02$ to $< 0.01$ \citep[][]{USERO15,BIGIEL16,GALLAGHER18}.

To harness the power of mm-wave spectroscopy to trace density across galaxies, we need an instrument that simultaneously achieves high spatial resolution, very high surface brightness sensitivity, and large spectral coverage (to capture many transitions at once). The ngVLA is exactly this instrument. The reference design review version of the ngVLA achieves rms sensitivity $0.01$~K per 10~km~s$^{-1}$ channel at $1''$ resolution in $1$~hour. The beam size of $1''$ matches the sizes of individual molecular clouds for galaxies out to the Virgo Cluster \citep[e.g., see][]{SUN18}.  At this resolution and for a characteristic cloud line width of $10$~km~s$^{-1}$, $\sim 0.01$~K is a typical brightness for HCN in a star-forming galaxy. Finally, the ngVLA bandwidth simultaneously captures almost all of the density-sensitive diagnostics in the $\lambda = 2.6{-}4$\,mm window  (depending on the exact distribution of the 20~GHz of bandwidth across the band), allowing a stunning gain in efficiency to carry out this fundamentally multi-line science. Previous telescopes require many tunings to cover the suite of lines seen in Figure \ref{fig:emis}.

\begin{figure}[t!]
\includegraphics[width=\textwidth]{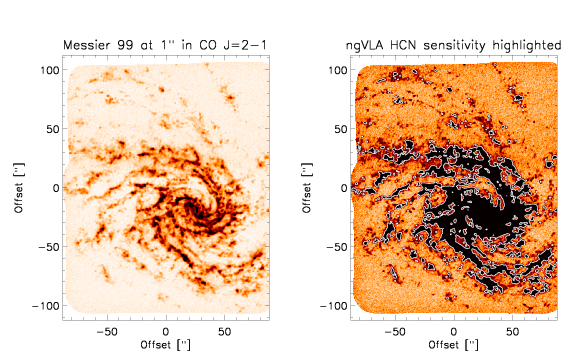}
\caption{{\bf The ngVLA Will Spectroscopically Image Faint Lines at High Resolution With Huge Multiplexing Capability.} CO~(2-1) peak temperature map (in a 10~km~s$^{-1}$ channel) for a local (Virgo cluster) star forming spiral at $1''$ resolution, sufficient to distinguish individual clouds, spiral arms, and distinct environments. On the right we show a version of the map scaled by a typical HCN-to-CO ratio \citep{USERO15} and saturated at the $1$~hour, $1''$, $10$~km~s$^{-1}$ line sensitivity of the ngVLA reference design, which is also shown by a white contour. The exercise shows that {\em the ngVLA will be able to detect high critical density lines on the scale of individual clouds in a few hours of integration. While doing so, it simultaneously observes a huge part of the cm- and mm-wave spectrum and covers a large area.} These capabilities make the ngVLA a density-mapping machine unmatched by any current or planned facility. In the text, we sketch how a $\sim 100$~hour program could be expected to measure the density distribution in individual molecular clouds across the whole active part of the galaxy.
\label{fig:map}}
\end{figure}

Put more simply, Figure \ref{fig:map} shows a $\sim 1''$ resolution CO~(2-1) map of a nearby star-forming spiral galaxy, M\,99 (NGC\,4254), from ALMA's PHANGS survey \citep[see][ data from A. K. Leroy, E. Schinnerer et al. 2019, in preparation]{SUN18}. In the right panel, we scale the CO map by a typical HCN-to-CO ratio and saturate the color scale at the rms line sensitivity achieved in one hour of integration (this level is also shown by a white line). This exercise shows that the fiducial sensitivity of the ngVLA in an hour is of the right order to detect faint, high critical density tracers. Integrations of a few hours or ten hours towards a part of a local galaxy should recover a large suite of faint lines in each beam, allowing the reconstruction of cloud-by-cloud density distribution across galaxies. 

To be more concrete, the ngVLA at 90~GHz could cover the inner, gas rich part of M99 (the illustrated galaxy in Figure \ref{fig:map}) in about 10 pointings. With an integration time of $\sim 10$~hours per pointing and $10$ pointings, a survey of the active part of this galaxy would cost $\sim 100$~hours. Such a survey would have a sensitivity of $\sim 0.003$~K per 10~km~s$^{-1}$ channel at physical resolution $\sim 75$~pc, covering at least half a dozen lines with different effective critical densities (CN, HCN, HCO$^+$, HNC, CS, their isotopologues, etc.). This would allow one to detect the dense gas in an individual molecular cloud ($M \sim 5 \times 10^5\,M_\odot$) at reasonable signal to noise down to dense gas fractions of a few percent in multiple tracers. This cloud-by-cloud view of the density distribution across a whole galaxy is exactly the kind of observation needed to address the key questions outlined above. It would be the kind of leap forward for dense gas that the pioneering PAWS survey \citep{SCHINNERER13} made for molecular gas: the first simultaneously sensitive, wide area, and high resolution view of gas (here dense gas, for PAWS CO emission) across a whole large part of galaxy.

\section{Complementary Science --- Chemistry and Physical Conditions Across Galaxies}

Density is not the only factor at play in the cm- and mm- part of the spectrum. Chemistry depends on environment and varies the abundance of each species as a function of environment \citep[e.g., see][]{VANDISHOECK14}. Both gas excitation and radiative transfer through clouds can play key roles in what we observe. A strength of the ngVLA's large proposed bandwidth is the ability to observe many transitions in one observation. Contrasting results from lines with similar critical densities will allow us to isolate the effects of density and chemistry. In addition, each density-probing observation will also be a line survey with potential sensitivity to the presence of shocks, the relative contributions of PDRs, XDRs, and CRDRs to that part of the galaxy, and excitation in the gas. Though we highlight density as a goal, this direction is inextricably linked to astrochemistry and the broader topic of physical conditions and excitation in the interstellar medium.

\section{Interface with Existing and Planned Facilities}

Many telescopes are making critical contributions to this topic right now including ALMA, NOEMA, the GBT, IRAM 30m, Nobeyama 45m, and the LMT. While no existing or planned facility approaches the power of the proposed ngVLA at the key frequencies $\nu < 116$~GHz, these other facilities are likely to do key work over the next decade that will inform future science directions with the ngVLA and have strong synergies with the ngVLA when it comes online.

Currently, ALMA is the best instrument available to the density-mapping experiments described above (soon to be paired with NOEMA). We expect that the state of the field when ngVLA comes online will likely be set by these two facilities. However, ALMA necessarily works at coarser resolution with less sensitivity and less bandwidth to map these lines. For matched resolution ($1''$), ALMA achieves $\sim 9$ times worse sensitivity in an hour with only a slightly bigger field of view. This calculation means that to match the sensitivity achieved by the ngVLA in $1$ hour, ALMA would have to integrate almost 100 hours, rendering large surveys impossible. ALMA also covers only $8$~GHz of bandwidth (though this is likely to be improved by upgrades). The ngVLA's proposed 20~GHz instantaneous bandwidth would allow vastly improved line multiplexing, covering a large part of the spectrum in a single observation. The sensitivity and bandwidth of the ngVLA will allow us to survey many lines in many different types of systems with relatively high resolution. The ngVLA promises a similar upgrade over the full NOEMA, which offers a wide ($16$~GHz) instantaneous bandwidth that allows for great multiplexing, but cannot match the raw sensitivity of the ngVLA due to the vast mismatch in collecting area.

Even so, ALMA and NOEMA will certainly blaze the trail towards the kind of science described above. Even with the ngVLA, ALMA will remain the key tool to study excitation, including the interface of excitation and density (likely by contrasting with ngVLA studies). This ability to combine high-$J$ (with ALMA) and low-$J$ (with ngVLA) observations will open an extremely powerful window into the physical state of gas in galaxies.

Similarly, the Green Bank Telescope, the IRAM 30-m telescope, the Nobeyama 45-m telescope, and the Large Millimeter Telescope will all make key contributions in this area over the next years. These single dish telescopes, however, necessarily emphasize wide field mapping, leveraging their surface brightness sensitivity, but do not resolve galaxies into individual clouds or physically distinct regions. These telescopes, especially if upgraded with new instrumentation, are likely to dramatically improve our integrated inventory of dense gas across the universe. These results will put the ngVLA in the position to resolve the physics related to density in discrete parts of galaxies: clouds, arms, and galaxy nuclei.

In thinking about the interplay of these facilities, a key point to keep in mind is that the stunning combination of bandwidth and collecting area for the ngVLA is guaranteed to open new windows into the physical state of gas across galaxies. Even in the most optimistic scenario, in which HCN and HCO$^+$ are mapped across many galaxies over the next ten years by the GBT, NOEMA, and ALMA, a suite of more sophisticated diagnostics --- optically thin isotopologues, chemicals that appear when CO freezes out, etc. --- lie just out of the most optimistic reach of current telescopes.  The ngVLA will be an incredible tool to access these lines.

\acknowledgements We thank the referee of this article, Fabian Walter, for helpful comments that improved the content. AKL gratefully acknowledges partial support by the National Science Foundation under Grants No. 1615105, 1615109, and 1653300.




\end{document}